\documentstyle[preprint,aps]{revtex}

\begin{document}
\draft
\preprint{IMSc/99/06/24}
\title{Anyons: Pseudo-integrability, Symmetry reduction and Semiclassical Spectrum}
\author{G. Date}
\address{The Institute of Mathematical Sciences,
CIT Campus, Chennai-600 113, INDIA.\\}

\maketitle
\begin{abstract} 

At the classical level anyons with harmonic confinement are known to exhibit 
two important properties namely partial separability and pseudo-integrability. 
These stem from the fact that this system is locally identical to isotropic
oscillator system but differs in the global topology of the phase space. We clarify
the meaning of pseudo-integrability and show that it amounts to a definite
reduction of the symmetry group. We elaborate on the role of the fundamental
group of the phase space and pseudo-intrgrability in the context of periodic
orbit theory and obtain evidence of non-exactly known eigenvalues from
the semiclassical trace formula. We also discuss an ambiguity regarding the
`half period' trajectories suggested by classical modeling and exhibited by
the exactly known propagator for two anyons. \\ 

\end{abstract}


\narrowtext

\section{Introduction}
Anyons as a research area is now over twenty years old \cite{review}. These
systems have emerged as being interesting in their own right from
mathematical physics point of view both at the classical and the quantum level. 
These systems constitute an example of inequivalent quantizations due to non
trivial fundamental group of the configuration space \cite{sigma}. These
two dimensional systems have a configuration space whose fundamental
group is a braid group \cite{braid}. Its one dimensional unitary 
representations give the kinematic classification labelled by the so 
called ``statistical" parameter $\alpha$. Being a system of finitely many 
degrees of freedom these are simpler to analyze for the dynamical 
consequences such as the spectrum of the Hamiltonian. \\

The quantum mechanical spectrum (with harmonic
potential added) shows two distinct qualitative features: a) eigenvalues values
which depend {\em{linearly}} on the statistical parameter, $\alpha$, all of 
which are exactly known and b) eigenvalues which depend {\em{non-linearly}} on
$\alpha$ and none of which is exactly known. These features were traced to 
the properties of partial separability and pseudo-integrability manifested 
at the classical level \cite{pseudo}. \\

Partial separability of the Hamiltonian was exhibited explicitly in terms 
of two collective degrees of freedom and the remaining ``relative" degrees 
of freedom. The total Hamiltonian, after removing the center of mass
degrees, was obtained as a sum of $H_1(collective) + H_2(collective, relative)$.
The commutator of $H_1$ and $H_2$ was shown to be proportional to
$H_2$. This implies that the subspace of eigenstates states of the full $H$ 
on which $H_2$ vanishes are exact eigenstates states of $H_1$ and these give 
{\em{all}} the exactly known eigenvalues values and eigenstates states. \\

Pseudo-integrability, a concept first introduced by Richens and Berry 
\cite{berry}, was described by exhibiting 2N constants of motion in
involution which fail to lead to integrability via usual action-angle
coordinates. This was conjectured to be the reason for the level repulsion
seen numerically in the non-linearly interpolating spectrum
\cite{pseudo}. The precise meaning of pseudo-integrability in this context, 
however, was not elaborated. In particular, how this classical feature 
translates into there being only two good quantum numbers, was not analysed. \\

There is another approach one could follow to get a handle on the many anyon
spectrum namely the stationary phase approximation (SPA) to the propagator 
$G(E + i \epsilon)$ developed using a suitable path integral representation
\cite{semiclassical}. In this approach the propagator is typically obtained 
as a sum over (families of) periodic trajectories in the classical phase space. 
This is entirely given in terms of classical quantities. One therefore expects 
to see directly the effects of non-trivial fundamental group of the phase space, 
pseudo-integrability and partial separability in a semiclassical framework. 
Since at the classical level, anyons with oscillator confinement, are 
{\it locally} identical to the oscillator system (but differ in the global 
topology of the phase space), the periodic orbits are known. Thus application 
of the periodic orbit theory (POT) to anyons, may be expected to be 
tractable.\\

There is one point to be noted at the outset. At the quantum level, the
statistical parameter, $\alpha_q$, enters via the stipulation of multivaluedness
of the wave functions and is {\it dimensionless}. At the classical level its
counterpart, $\alpha_c$, enters as the coefficient of a total derivative
term in the Lagrangian and has the dimensions of $\hbar$. These two must be
related as $\alpha_c = \hbar \alpha_q$. In the SPA computation, $\alpha_c$ is
held fixed with $\hbar$ going to zero. In the leading approximation, the
spectrum will depend {\it linearly} on $\alpha_c$ and thus also on $\alpha_q$. 
Alternatively, for comparison with the quantum spectrum to leading order, 
one will want to keep $\alpha_q$ fixed implying that $\alpha_c$ to go to zero
with $\hbar$. Viewed either way, one may {\it not} expect to see 
{\it non-linear} dependence on the statistical parameter at the level of
the leading approximation. Nonetheless, if the non-linearly interpolating 
eigenvalues also have a linear piece, one should be able to see it in the 
leading SPA. \\

In this work, we demonstrate that the classical analysis is adequate for
exhibiting quantum symmetries. In this process we sharpen and clarify the
meaning of pseudo-integrability and role of fundamental group of the phase
space. \\

We also sketch application of the periodic orbit theory to many anyons.
In the process we show the presence of eigenvalues which are potentially
non-linearly interpolating. Such eigenvalues have been seen in the numerical
spectrum for three and four anyons \cite{sub-numerical} and have been 
conjectured to be present for general $N$ \cite{numerical}. We obtain these 
systematically from POT. \\

Further, both the classical modeling and the exact propagator for two anyons,
indicate an ambiguity regarding possibility of closed orbits with half the
basic period and its inclusion in the trace formula. We discuss this ambiguity 
in some details. \\

The paper is organized as follows. \\

In section II we specify the classical model. The system is shown to be 
identical to just the isotropic oscillator {\em{locally}} but not at 
the global topological level. The set of classical trajectories is detailed.  \\

In section III we  classify the classical trajectories using symmetry
transformations and explain how the non-trivial global topology reduces the
symmetry group. \\

In section IV we sketch the POT application to anyons and demonstrate
existence of eigenvalues other than the exactly known ones. \\

Section V contains a discussion of the issue of half period trajectories. We
show that it is possible to regularise the point-like statistical flux and
deduce the existence of half period trajectories. We look for their evidence
in the exact propagator for two anyons and point out their ambiguous status. \\

Section VI contains general remarks on the role of the fundamental group,
possible reduction of symmetries due to non-trivial topology of the phase
space. A summary of results and conclusions is presented. \\

Appendix A contains details of $OSp(4N, R)$ symmetry. \\

Appendix B contains discussion of regularised classical dynamics of anyons 
and the presence of half period trajectories. \\

\section{Classical modeling and dynamical trajectories }

Anyons are fundamentally defined as two dimensional quantum mechanical
systems with a specified multi-valuedness for its wave function. The
multi-valuedness is stipulated in terms of one dimensional non-trivial 
representations of the fundamental group of the configuration space, $Q$. 
At the least one defines the configuration space for N anyons to be,

\begin{equation}
Q \sim R^{2N} - \Delta , 
\end{equation}

where $\Delta$ is the set of points in the plane  at which two or more
particle positions coincide. Denoting by $\vec{r}_{ij}$ the difference
of $\vec{r}_i$ and $\vec{r}_j$ ,

\begin{equation}
\Delta \equiv \{ \vec{r}_i ~ / ~  \vec{r}_{ij} = \vec{0} ~ \mbox{for some} ~ i \neq j \}
\end{equation}

The fundamental group of $Q$ is the so called pure braid group and is known
to be non-trivial \cite{braid}. ( For indistinguishable particles $Q$ should be
modded by the permutation group $S_N$. For most of what follows, this is
not essential. ) The multi-valued wave functions can be
expressed in terms of single valued wave functions by a specific explicit
multi-valued pre-factor while differentials of multi-valued wave functions 
can be expressed as ``covariant" differentials of single valued wave functions.
This is discussed in detail in reference \cite{dmm}. Using the same notation
as ref. \cite{dmm}, in terms of single valued wave functions, the quantum 
mechanical Hamiltonian for N-anyons with harmonic confinement is given by,

\begin{equation}
H = \frac{1}{2}\hbar\omega \{ \sum_i ( \vec{p}^2_i + \vec{r}^2_i ) - 
\alpha_{q} \sum_{i \neq j} \frac{\vec{r}_{ij} \times \vec{p}_{ij}}{r_{ij}^2} + 
\alpha_{q}^2 \sum_{i \neq j,k} \frac{\vec{r}_{ij}\cdot\vec{r}_{ik}}{r^2_{ij} r^2_{ik}}
\}
\end{equation}

Here $\vec{p}_i ~ \equiv ~ -i \vec{\nabla} $. It is convenient to introduce 
the ``statistical" gauge potential $\vec{A}_i$ as,

\begin{equation}
\vec{A}_i ~ \equiv ~ \alpha_{q} \sum_{j \neq i} \hat{k} \times 
\frac{\vec{r}_{ij}}{r^2_{ij}}
\end{equation}

where $\hat{k}$ is the unit vector in the z direction. Introducing
$\vec{\pi}_i ~ \equiv ~ \vec{p}_i ~ - ~ \vec{A}_i $ , the Hamiltonian 
can be expressed as ,

\begin{equation}
H =  \frac{1}{2}\hbar \omega \{ \sum_i ( \vec{\pi}^2_i + \vec{r}^2_i ) \} .
\end{equation}

It is easy to see that the $\vec{\pi}_i$ and the $\vec{r}_i$ satisfy the same
commutation relations as $\vec{p}_i$ and $\vec{r}_i$ away from the set
$\Delta$. \\

{\bf Remarks:}

{\bf 1)} In the above quantum expressions $\vec{p}_i, \vec{r}_i, 
\alpha_{q}$ are all dimensionless. For the classical expressions below we will
use $\alpha_{c}$ which has the dimensions of $\hbar$. These two are related by 
$\alpha_{c} \equiv \hbar \alpha_{q}$.

{\bf 2)} The use of  $\vec{\pi}_i , \vec{r}_i$ coordinates is the classical
counter part of the quantum ``anyonic gauge" where the Hamiltonian operator
expression looks simple and the wave functions are multi-valued. \\

The corresponding classical system is defined with $Q$ as the configuration
space and its cotangent bundle as the phase space $(\Gamma)$. The Hamiltonian 
is the same as one given in the eq.(3) above. Noting
that the Poisson Brackets (PB) among $\vec{p}_i, \vec{r}_i$ are the same as
those among $\vec{\pi}_i , \vec{r}_i $ we see that going from $\vec{p}_i$ to
$\vec{\pi}_i$ is a canonical transformation. The classical Hamiltonian may
then be taken to be as given in eq.(5). {\it Locally}  then the anyon 
system is identical to the 2N dimensional isotropic oscillator. The only 
difference is in the global topology of the phase space which turns out to 
have non-trivial consequence. \\

The Lagrangian from which the Hamiltonian, eq.(3), may be obtained is given by,

\begin{equation}
L = \frac{1}{2} \left\{ \sum_i (~ \dot{\vec{r}}_i^2 - \vec{r}_i^2 ) + \alpha_{c} \sum_{i \neq j} \dot{\theta}_{ij} ~) \right\} , 
~~~~ \theta_{ij} \equiv tan^{-1}( \frac{y_{ij}}{x_{ij}}) .
\end{equation}

The $\alpha_{c}$ dependent term being a total derivative implies that the
equations of motion are identical to those of the oscillator. The class of
solutions is of course {\it not} the same due to the removal of the set
$\Delta$. Note also that there is no explicit appearance of $\alpha_{c}$ either
in the equations of motion or their solutions. Presence of $\alpha_{c}$ only
changes the classical action and also affects the set of solutions
qualitatively. \\

The configuration space equations of motion are:

\begin{equation}
\ddot{\vec{r}}_i ~~ = ~~ \vec{r}_i ~~~ \forall i ~~~~ \Rightarrow ~~ \ddot{\vec{r}}_{ij} ~~ = ~~ \vec{r}_{ij}.
\end{equation}

Therefore in the plane defined by any two vectors $\vec{r}_i$ and $\vec{r}_j$
the difference vector traces an ellipse in general. For the oscillator
degenerate ellipse (straight line) is permitted but not for the anyons.\\

Looking at the Hamiltonian form, one may expect to have the same symmetries
for anyons as for the oscillator, namely the full OSp(4N,R) symmetry. Indeed
rank of this group being 2N implies existence of 2N constants of motion in
involution. These were exhibited in reference \cite{pseudo} and were stated to be
indicative of the pseudo-integrability property. It was pointed out that
integrability via action-angle coordinates requires further global input. 

\section{Classification of orbits and symmetry reduction}

In this section we will carry out a classification of orbits for $N$ anyons
and see how the pseudo-integrability amounts to a reduction of the dynamical
symmetry. We will also see the role played by the fundamental group of the
phase space in this regard. Noting that a trajectory is specified by giving
an initial point, we will use the dynamical symmetry of the oscillator system
to group various orbits into continuous families. For anyons, certain points
and hence certain orbits are disallowed (removed) with the result that one
gets several continuous families. 

\subsection{The case of two anyons}

As usual we can separate the center-of-mass (CM) dynamics from the relative
coordinate dynamics. The CM dynamics is trivial -- identical to an
oscillator locally and globally. The anyonic feature is contained entirely in
the relative dynamics described as:

\begin{equation}
\begin{array}{lclclcl}
Q & = & R^2 - \vec{0} &~~~ ,~~~ & \Gamma & = & Q \times R^2 \\
L & = & \frac{1}{2} ( \dot{\vec{r}}^2 - \vec{r}^2 ) + \alpha_{c} \dot{\theta} &
~~~,~~~ & \theta & \equiv & tan^{-1}{y/x} \\
H & = & \frac{1}{2}(\vec{\pi}^2 + \vec{r}^2) &~~~ ,~~~ & \vec{\pi} & \equiv & 
\vec{p} - \alpha_c \frac{\hat{k} \times \vec{r}}{r^2}  
\end{array}
\end{equation}

The classical trajectories are trivially known in these cases. These are
ellipses with two constants of motion, the energy ($E$) and the angular 
momentum ($J \ = \ (\vec{r} \times \dot{\vec{r}})_z $). The sign of $J$ gives
the sense of traversal and a trajectory reaches the point $\vec{r} \ = \ \vec{0}$
if and only if $J \ = \ 0$. These degenerate trajectories ($J \ = \ 0$) are
not allowed for the anyons though are allowed for the oscillator. \\

Consider the set of all possible trajectories with a given energy. Fix two 
such trajectories with angular momenta $J, J^{\prime}$. Suppose there exists a
continuous family of interpolating trajectories connecting these two. Clearly
if the angular momenta have opposite signs, then at least one of the
interpolating trajectories must be degenerate. But this is disallowed for
anyons. Hence, for anyons there must be {\it {at least}} TWO `orientation' classes 
of trajectories distinguished only by the sign of angular momenta. \\

While we have shown that there must be at least two families we do not know 
yet if there {\it {must}} be {\it {precisely}} two families. We will use the 
symmetries to show that there are precisely two families for anyons and 
precisely one family for the oscillator. \\

As far as the PB algebra is concerned, the oscillator and anyons
are identical and hence we do have {\it {infinitesimal}} symmetries forming the
Lie algebra of $U(2)$ (or $OSp(4,R)$) for both cases. The constant $H$
surface ($S^3$ in the phase space $R^4$ for oscillator) is of course
invariant under the infinitesimal symmetry transformations. For anyons, the
phase space has $\vec{r} = \vec{0}$ removed and the constant $H$ surface has
a great circle (one dimensional) removed from the $S^3$, say a ``longitude". \\

In the appendix A we have given the details of the $OSp(4N,R)$ symmetry of the
oscillator. In the present case of two anyons with center of mass coordinate
separated, we have effectively, $N \ = \ 1$ and the symmetry generators form
$OSp(4, R)$. We have only the 4 generators of the ${\bf {T}}_i$ type.
Of these four, ${\bf u}_1$ generates rotations of position and
momentum while ${\bf u}_2$ generates time evolution. The generic integral 
curves are given in the appendix A. \\

From appendix A we also recall that $OSp(4, R)$ acts transitively on $H = E$ 
sphere.
This shows that for the oscillator there is precisely ONE family of
trajectories. For anyons we need analogous result with the extra condition
that degenerate trajectories are avoided to conclude that there are precisely 
TWO families. \\

{\underline {Claim}} Given any two trajectories with the same sign of their
angular momenta, there exist a one parameter group connecting the two without
changing the sign of angular momenta. \\

{\underline {Proof}} Observe that ${\bf u}_1, {\bf u}_2$ transformations
leave the angular momentum invariant. Choose ${\bf u} \ = \ cos(\beta){\bf
{u}}_3 \ + \ sin(\beta){\bf {u}}_4 $. The one parameter group generated by
this generator transforms the angular momentum as, (appendix A)

\begin{equation}
2J(\sigma) ~ = ~ 2J -sin^2(\sigma)\left\{ 2J - \bar{{\bf {\omega}}}\bar{ {\bf u}} 
{\bf {L u \omega}} \right\} + sin(2 \sigma) \left\{ \bar{ {\bf {\omega }}} {\bf { L u
\omega}} \right\}
\end{equation}

For the choice of {\bf u} made, $\bar{ {\bf u}} {\bf {L u}} \ = \ - {\bf L} $
and putting $J_u \ \equiv \ \frac{ \bar{ {\bf {\omega }}} {\bf { L u \omega}} }{2}
$ ,

\begin{eqnarray}
J(\sigma) & ~ = ~ & J  \ cos(2 \sigma) + sin(2 \sigma)  \ J_u \nonumber \\
& ~ = ~ & \sqrt{J^2 + J_u^2}  \ cos( 2 \sigma - \delta ) ~~, ~~~~~ \delta ~ 
\equiv ~ tan^{-1}(J_u/J) ~  \in ~ (-\pi /2 , \pi /2)
\end{eqnarray}

This is valid for all choices of $\beta$. Observe that $J_u$ depends on
$\beta$. Furthermore it satisfies,

\begin{eqnarray}
\frac{\partial^2 J_u(\beta)}{\partial \beta^2} ~ & = & ~ - J_u(\beta) 
~~~~~~~~~~~~~~~~~~~~~~~~~~~~~~~~~ \Rightarrow \nonumber \\
J_u(\beta) ~ & = & ~ J_u(0) cos(\beta) + \frac{\partial J_u}{\partial \beta}(0)
sin(\beta) ~~~~~~~~ \Rightarrow \\
(J_u)_{max} ~ & = & ~ \ \left\{ \ (J_u(0))^2 \ + \ \right.
\left. ( \frac{\partial J_u}{\partial \beta}(0) )^2 \ \right\}^{1/2} 
~~~~~ = ~~ \sqrt{ \ E^2 \ - \ J^2 \ } \nonumber
\end{eqnarray}

The last equality follows by explicit evaluation. Choosing $\beta$ to
maximise $J_u$ then implies that the square root in the equation for 
$J(\sigma)$ is just the energy $E$. Since for any given energy, we must have 
$J^2 \le E^2$, it follows that $J(\sigma)$ covers the entire range of possible 
$J$ values. \\

From this it follows that we can always choose a $\hat{\sigma}$ such 
that $J(\hat{\sigma}) \ = \ J^{\prime}$ without passing through zero angular
momentum. This proves the claim. Hence there are precisely TWO families of 
trajectories each spanning the whole orientation class. This gives the
classification of trajectories.\\

One may view the $S^3$ as compactification of $R^3$ with the south pole as the
origin of $R^3$ and the north pole as the point at infinity. The removal of
$\Delta$ now corresponds to removal of an infinite line, say, the z-axis of
the $R^3$. It is immediate that the fundamental group is the group of
integers, $Z$, and therefore the basic trajectories winding around the z-axis 
belong to two orientation classes. In this case the fundamental group directly 
shows how the basic trajectories are naturally classified into the ``orientation 
classes" . \\

Consider now the Hamiltonian vector fields generating the three one parameter 
subgroups of $SU(2), {\bf u}_1$ generates rotations
about the z-axis. Its orbit through any generic point avoids the
z-axis and the vector field remains complete (its integral curves range over the
full real line) . The other two subgroups on the other hand have orbits 
{\it {necessarily}} cutting through the z-axis. These vector fields therefore
are necessarily incomplete (for the anyon case). The transformations generated by
these vector fields do {\it{ not}} for a group and the dynamical symmetry group for
anyons is reduced from $U(2)$ to $U(1) \times U(1)$.\\

Thus the removal of the set $\Delta$ has altered the fundamental group, has
classified trajectories into orientation classes and has also reduced the
dynamical symmetry group form $U(2)$ to $U(1) \times U(1)$ due to the
incompleteness of vector fields. \\

Note that while the symmetry group is reduces from $U(2)$ to $U(1) \times
U(1)$ for anyons, the rank has remained the same and therefore the anyonic system 
continues to be integrable. The integral curves being circles, we have
integrability via action-angle variables and we get the exact spectrum from 
the semiclassical approximation. In fact one can follows the EBK quantization
procedure to reproduce the exact spectrum. \cite{dmrev}. 

\subsection{The case of many anyons}

The $N > 2$ case differs from the $N = 2$ case in many ways. The potential
symmetry group, $OSP(4N, R)$, (see appendix A) generating families of trajectories 
is lot more complicated and so is the fundamental group for the phase space of 
anyons. However the analysis of the previous section already suggests a strategy 
for obtaining a classification of the families of trajectories. A generic 
trajectory may be viewed as a collection of $\frac{N ( N - 1 )}{2}$ ellipses 
traced by the $\vec{r}_{ij} , i < j $. To get a convenient handle on the 
disallowed trajectories, define
$J_{ij} \ \equiv \ ( \vec{r}_{ij} \times \dot{\vec{r}}_{ij} )_z $. Clearly a
trajectory will cut through the set $\Delta$ if and only if $J_{ij}$ vanishes
for at least one pair of indices. We will term such a trajectory as degenerate. 
For anyon only non-degenerate trajectories are allowed. Note that $J_{ij}$
are constants of motion just as $J_i$ and $E_i \equiv \frac{1}{2}( r_i^2 +
p_i^2)$ \cite{pseudo} are. Each of these elliptical trajectories will have 
a sense of traversal ( $ \epsilon_{ij} \ \equiv \ $ sign of $ J_{ij}$ ). Thus
a basic trajectory of $N$ particles has an associated set: \{$\epsilon_{ij}$\}.
Exactly as in the previous sub-section, the set of non-degenerate trajectories 
can be classified by these `orientations'. Provided we can exhibit trajectories 
which realise all possible choices of $\epsilon_{ij}$, we will have 
{\it at least} $2^{N ( N -1 )/2}$ families for anyons and precisely 
as many if every member of a class is connected to every other member of the 
same class by a symmetry transformation without crossing $\Delta$. It is easy
to show that every choice of $\epsilon_{ij}$ is realised by considering
concentric circular trajectories for each of the particles with various
possible signs for the angular momenta $J_i$ and various possible ordering of
the radii of the circular trajectories. For oscillator we will have precisely 
one family if $OSp(4N, R)$ acts transitively on the constant energy sphere, 
$S^{4N -1}$. \\

The proof of transitivity in the case of oscillator is most directly obtained
by noting that $OSp(4N, R)$ is isomorphic to $U(2N, C)$ and it is well known
(and easy to see) that the coset space $U(2N,C)/U(2N -1, C)$ is diffeomorphic
to $S^{4N -1}$. This immediately shows that for the oscillator, there is
precisely ONE family of periodic trajectories.  The counting of families is
more tricky in the case of the anyons. Let us see the symmetry reduction in a
slightly different manner.\\

Let $\Delta^{\prime}$ denote the set of all degenerate trajectories i.e. at
least one of the $J_{ij}$'s is zero. Note that $\Delta$ and $\Delta^{\prime}$
are different, the former being a proper subset of the latter. Suppose we find
the subgroup of symmetries which leave $\Delta^{\prime}$ invariant, then the
same subgroup will also leave the set of all non-degenerate trajectories
invariant. Since we are looking for a group, we can consider the
infinitesimal action. Let $J \ \equiv \ \prod_{i < j} J_{ij}$. Clearly, $J =
0$ characterises the set $\Delta^{\prime}$. If we have a degenerate
trajectory with two or more $J_{ij}$'s zero, then every infinitesimal action
will keep within the set of degenerate trajectories. If only one $J_{mn}$ is
zero then infinitesimal action will act invariantly if and only if $J_{mn}$
is invariant. Of course we could have degenerate trajectories with different
angular momenta being zero and hence for the invariance of $\Delta^{\prime}$ it is
necessary that the generators of the subgroup must leave each of the $J_{ij}$
's invariant {\it whenever $J_{ij}$ itself is zero}. This subgroup is
identified in the appendix A. There are two forms of generators. Any
generator with all the block matrices being the same {\bf u} matrix with
${\bf u} \ \in \ OSp(4,R)$ will leave all the J's invariant. 
These transformations act on the center-of-mass variable alone and
constitute the expected $OSp(4,R)$ symmetry. In addition, generators with only
diagonal blocks being the same {\bf u} and all the off-diagonal blocks being
{\bf 0} also leaves the J's invariant provided ${\bf {u \ = \ u}_1}$ or ${\bf
u}_2$ (or a linear combination thereof). These are just the total Hamiltonian 
and the total angular momentum ($\sum_i J_i$) which generate time evolutions and 
common rotations of all the positions and momenta. These are the only subgroups 
leaving the set of degenerate (and non-degenerate) trajectories invariant. 
The vector fields corresponding only to these will remain complete. Thus we see 
that the symmetry is reduced to $OSp(4,R) \times O(2) \times O(2)$. This is 
identical to the symmetry in the case of two anyons.\\

It is apparant from the above discussion that pseudo-integrability of anyons
really means that not all the Lie algebra level symmetries exponetiate to
Lie group symmetries. One should distinguish between ``infintesimal
integrability" and integrability. A general discussion is given in the last
section. \\

This purely classical analysis has already explained the qualitative result that
for anyons only the total energy and the total angular momentum are the
conserved quantities (good quantum numbers). Since we also have a
classification of periodic trajectories we will take the next step of
attempting semiclassical approximation to see what further information can be
obtained. \\

\section{Semiclassical Spectrum for anyons}

The central quantity of interest is the propagator defined as,

\begin{eqnarray}
G(E + i\epsilon) & \equiv & Tr( \frac{1}{E - H + i \epsilon} ) \nonumber \\
& = & \sum_{n} (E - E_n + i\epsilon)^{-1} \\
& = & \sum_{n}  ~ {\cal P} ( (E - E_n)^{-1} )  - i \pi \sum_{n} ~\delta (E - E_n) \nonumber 
\end{eqnarray}

In the last equality, ${\cal P}$ denotes the principle value while the second term is
of course the density of states. Defining $\tilde{G} \equiv i G$ allows us
to write the density of states, $g(E)$ as ,

\begin{eqnarray}
g(E) & = & \frac{1}{2\pi} ~ \{ \tilde{G}(E + i\epsilon) + \tilde{G}^{*}(E - i \epsilon)
~\} \nonumber \\
 & = & \frac{1}{\pi} ~ Re(~ \tilde{G}(E + i\epsilon) ~)
\end{eqnarray}

The derivation of the semiclassical trace formula begins with the definition
of the propagator, $G(E)$, with trace operation expressed via a path integral
representation. The trace operation combined with the stationary phase
approximation (SPA) gives the propagator as a sum over periodic orbits (in the
phase space), of terms with an amplitude given by a suitable Van-Vleck determinant
times a phase whose exponent is the value of the `action', $\int {\bf {p\cdot
dq}}$ around the periodic orbit together with a contribution from the Maslov
indices. This works very well for isolated periodic orbits. However for
systems with symmetries, as for the oscillator or anyons, the periodic orbits
come in continuous families and a modification is needed. \\

In such cases, the sum over orbits gets replaced by a sum over families of
orbits together with a measure factor. If the family arises due to a symmetry,
as is usually the case, the amplitude and the phase is same
for all members of the family and as such can be computed from any one
member. This is discussed by Littlejohn et al \cite{littlejohn} in detail. 
The origin of families of periodic orbits in the case of oscillator is of course
the $OSp(4, R)$ dynamical symmetry and for anyons essentially the same 
transformations generate families of trajectories. This is already discussed
in the previous section.\\

To obtain the action along a periodic orbit one point needs to be noted.
Observe that in terms of the $\vec{\pi}_i , \vec{r}_i$ coordinates 
there is no explicit reference to $\alpha$ and $\Delta$ is also precisely the 
set on which one gets $\delta$-function singular PBs with $\alpha$ appearing as 
the coefficient. If one did inverse Legendre transformation to get a Lagrangian
one would not get the $\alpha$ dependent term, but of course the configuration
space will have the $\Delta$ removed. Using $\vec{p}_i, \vec{r}_i$ variables
and then doing inverse Legendre transformation will produce the Lagrangian
with the $\alpha$ dependent total derivative term. Now the removal of
$\Delta$ is made explicit by the $\alpha$ dependent term. It is this explicit
form which is convenient for computation of action for periodic orbits.\\

The action for the oscillator is equal to $(2\pi E)/\omega$ and the action for 
anyons is the same as that for the oscillator except for the 
contribution from the total derivative term in the Lagrangian. Its value is 
$\pm 2\pi \alpha$ with sign depending on the orientation class(es). We have
already seen that for $N = 2$ we have precisely $2$ orientation classes while
for general $N$ we have {\it{at least}} $2^{N(N-1)/2}$ classes. \\

Observe that {\it {even without}} the knowledge of precise number of
families, we have precisely $2^{N(N - 1)/2}$ sums in the sum over
periodic trajectories. This is because, the contribution of the action
integral depends only on the orientation class and not on further possible
subclasses of trajectories. The measure factors therefore add up within each
orientation class and this is sufficient to get the semiclassical
eigenvalues. \\

The sum over families in the trace formula contains as many terms. To get the 
eigenvalues we need to look at only the action integral which differs from the 
oscillator only by the additional $\oint \frac{\alpha_c}{2} \ \sum_{i \ne j} 
\dot{\theta}_{ij}$ contribution. The oscillator's contribution to the action 
integral is of course just $\frac{2 \pi}{\omega} E$, where $E$ is the classical 
energy and $\omega$ is the oscillator frequency. \\

The net $\alpha_c$ dependent contribution to the action integral is given by,

\begin{equation}
\frac{\alpha_c}{2} \oint \sum_{i \ne j} \dot{\theta}_{ij} ~ = ~ 2 \pi \alpha_c \sum_{i
< j} \epsilon_{ij} 
\end{equation}

Here $\epsilon_{ij}$ is the sign of traversal of $\vec{r}_{ij}$ and is $\pm
1$. The $\sum_{i < j}$ is trivial to evaluate. \\

Since the classical trajectories are identical to those of the oscillator, we
will have the same Maslov indices which incorporate the usual zero point
energy namely, $N(N-1)$ for $2N$ dimensional oscillator. Including this
contribution, the semiclassical energies are given by :

\begin{equation}
E_n \{\epsilon_i\} ~ = ~ \hbar\omega (n ~ + ~ 
\alpha_q \sum_{i < j } \epsilon_{ij} ~ + ~ N(N-1) ~ )~~~~~~~~~~ n ~ \ge 0
\end{equation}

The coefficient of $\alpha_q$ takes all possible integer values from 
$ - N (N -1)/2 \ $ to $ \ N (N -1)/2 \ $ in steps of 2. The extreme
values correspond to all the particles traversing the same way and represent
the contribution of ``collective" motion shown in \cite{pseudo}. The
remaining values represent contributions of ``relative" motion. These are
new eigenvalues albeit at the leading $\hbar$ level. Numerical spectra
available for $N = 3, 4$ show eigenvalues matching with the above at the
$\alpha_q \ = \ 0$ and $\alpha_q \ = \ 1$ \cite{sub-numerical}. These numerical 
eigenvalues however are non-linearly interpolating while our semiclassical 
eigenvalues are linearly interpolating. It is possible that the semiclassical 
eigenvalues will receive corrections to make them non-linearly interpolating. 
In reference \cite{numerical} also it was conjectured that there will be
linearly interpolating eigenvalues with various slopes based on a different
semiclassical argument. We have obtained these eigenvalues by direct
application of the periodic orbit theory and proved their existence for all
$N$. \\

Observe that the $\alpha_q$ dependence in the eigenvalues ( locations of 
poles of $G(E)$ ) arises only from the explicit $\alpha_c \equiv \alpha_q \hbar$ in
the action. This therefore must be at the most linear in $\alpha_q$. Since
$\alpha_q$ is fixed, the semiclassical limit of $\hbar \rightarrow 0 $ 
implies $\alpha_c \rightarrow 0$. Thus one may at the most see linear
$\alpha_c$ dependence in the leading approximation which could however be
indicative of exact non-linearly interpolating eigenvalues. \\

Thus the leading level semiclassical propagator already indicates the
presence of eigenvalues such that $E(\alpha_q = 1) - E(\alpha_q = 0)$ takes
all possible integer values between $\pm N(N-1)/2~$ in steps of 2. \\

\section{Half period trajectories and exact propagator for two anyons.}

Recall that anyons are fundamentally defined quantum mechanically and all the
energy eigenfunctions of the system vanish on the set $\Delta$ of coincident points.
We modeled the classical system by removing $\Delta$ and explored the 
consequences in the previous sections. In particular we just {\it {omitted}}
trajectories that could cut through $\Delta$.\\

From a purely classical point of view, though, this is little unsatisfactory. 
One could legitimately ask just what happens to trajectories (eg $J = 0$ for $N = 2$)
that attempt reaching the disallowed region? To decide this one has to {\it{extend}}
the classical modeling by supplementing the equations of motions with a 
specified ``boundary" condition. The choice is helped by noting that the curl 
of the vector potential is a $\delta$-function which may be ``regulated" to 
stipulate the ``boundary condition". This analysis is given in appendix B. \\

The result is that {\it {all the elliptical trajectories are identical to 
those of the 
oscillator and only the degenerate trajectories get modified. A degenerate 
trajectory must 
reflect back from the coincident point.}} \\

Clearly such an trajectory will have half the period of the generic trajectories and hence 
we will refer to these as the ``half trajectories". \\
 
Should these orbits be included in the sum over orbits in the context of 
semiclassical approximation?  Are they already ``visible" in the exact spectrum 
known for two anyons? To answer these, let us look at the exact propagator
for the (relative coordinate part of) the two anyon system. \\

The propagator can be computed exactly from the known exact spectrum \cite{bhaduri}.
The exact spectrum is given by,

\begin{equation}
E_{n,j} ~ = ~ \hbar\omega ( 2n + | j - \alpha_{q} | + 1 ), ~~ n \ge 0,~~  j \in Z
\end{equation}

From this the partition function is obtained as,

\begin{equation}
Z(\beta) ~ = ~ \frac{cosh(~ \beta\hbar\omega~ (\alpha_{q} - 1)~ ) +
cosh(~ \beta\hbar\omega\alpha_{q}~ )}{2 sinh^2(\beta\hbar\omega)}
\end{equation}

and the exact density of states is obtained as:

\begin{eqnarray}
g(E) & \equiv & \frac{1}{2\pi i} ~ \int_{\epsilon - i \infty}^{\epsilon + i
\infty} e^{\beta E} Z(\beta)  \nonumber \\
& & \nonumber \\
& = & \frac{E}{(\hbar\omega)^2}\{ 1 + \sum_{k \ge 1}\{ 
cos(\frac{2\pi k E}{\hbar\omega} + 2\pi k \alpha_q ) + 
cos(\frac{2\pi k E }{\hbar\omega} - 2\pi k \alpha_q ) \} \nonumber \\
&  & - \frac{1}{(\hbar\omega)} \sum_{k \ge 1}\{ 
2\alpha_q ~  sin(2\pi k\alpha_q) ~  sin(\frac{2\pi k E }{\hbar\omega}) \} \\
&  & + \frac{1}{(\hbar\omega)} \sum_{k \ge 1}\{ 
(-1)^k  ~ sin(\pi k\alpha_q)  ~ sin(\frac{\pi k E }{\hbar\omega}) \} \nonumber 
\end{eqnarray}

Rewriting the products of sines as differences of cosines we get,

\begin{eqnarray}
g(E) & = & \frac{E}{(\hbar\omega)^2}  \nonumber \\
&  & + \frac{1}{\hbar\omega} ~ (~ \frac{E}{\hbar\omega} + \alpha_q ~)
\sum_{k \ge 1} ~ cos(\frac{2\pi k E}{\hbar\omega} + 2\pi k \alpha_q )  ~ 
\nonumber \\ 
&  & + \frac{1}{\hbar\omega} ~ (~ \frac{E}{\hbar\omega} - \alpha_q ~)
\sum_{k \ge 1} ~ cos(\frac{2\pi k E}{\hbar\omega} - 2\pi k \alpha_q )  ~ \\ 
&  & - \frac{1}{2\hbar\omega} ~ 
\sum_{k \ge 1} ~ (-1)^k ~ cos(\frac{\pi k E}{\hbar\omega} + \pi k \alpha_q ) 
\nonumber \\ 
&  & + \frac{1}{2\hbar\omega} ~
\sum_{k \ge 1} ~ (-1)^k ~ cos(\frac{\pi k E}{\hbar\omega} - \pi k \alpha_q ) 
\nonumber  
\end{eqnarray}

{\bf {Remarks:}}

1. The partition function is manifestly invariant under $\alpha_q \rightarrow
1 - \alpha_q$ . The convergence of integrals in computing the density of states for 
$\alpha_q \in (0, 1)$ requires $\frac{E}{\hbar \omega}$ to be greater than or 
equal to 1. 

2. The first term above is the usual Thomas-Fermi term \cite{bhaduri}.	This
will be suppressed in expressions below. 

3. In the limit $\hbar \rightarrow 0$ keeping $E, \alpha_q$ fixed, the exact 
propagator  effectively looses all dependence on $\alpha_q$. This is also the 
semiclassical trace formula for the oscillator showing that the semiclassical 
approximation is exact for the oscillator. All the other terms are sub-leading.

4. In the semiclassical computation, one will use $\alpha_c = \hbar \alpha_q$
and the SPA will be done keeping $\alpha_c$ fixed. Then only the last two
terms will be sub-leading and the propagator will continue to have $\alpha_c$
dependence. 

5. In the last two sub-leading terms the argument of cosine is half of that in 
second and the third term. As is well known, the argument of 
cosines is the classical action, $\int \vec{p}\cdot d\vec{r}$, around a classical 
orbit including possible contributions from ``Maslov indices". The last two terms
are therefore suggestive of contribution from ``half orbits". Indeed if one takes 
a Fourier transform of $g(E)$ \cite{bhaduri} , then one sees a peak at 1/2 period from the 
last two terms. {\it One may therefore conclude that half orbits should also be
included in the trace formula.} \\

This turns out to be somewhat ambiguous. To see this let us express the propagator 
in a sum-over-orbits form. To do this rewrite the the last two terms to resemble 
the second and the third terms. \\

The last two terms also have a $(-1)^k$ in the summation over $k$ and this can
be handled in two ways. \\

(A) Separate these sums into even and odd integer sums. All the sums over $k$ 
being geometric series can be done explicitly 
($i\epsilon$ needs to be added). Defining ${\cal{E}} = E/(\hbar\omega)$, we get,

\begin{eqnarray}
\hbar\omega g(E) & = & \cal E ~ + \nonumber \\
& & \frac{\frac{1}{2} ~ ({\cal E} + \alpha_q - \frac{1}{2} ~ )e^{2\pi i ( {\cal E}
+ \alpha_q )} + \frac{1}{4} e^{i\pi ({\cal E} + \alpha_q) } }{1 - e^{2\pi i \alpha_q
({\cal E} + \alpha_q )}} ~ + ~ \mbox{C. C.} \\
& & \frac{\frac{1}{2} ~ ({\cal E} - \alpha_q + \frac{1}{2} ~ )e^{2\pi i ( {\cal E}
- \alpha_q )} - \frac{1}{4} e^{i\pi ({\cal E} - \alpha_q) } }{1 - e^{2\pi i \alpha_q
({\cal E} - \alpha_q )}} ~ + ~ \mbox{C. C.} \nonumber 
\end{eqnarray}

Here C. C. means complex conjugation.  Now we can ``read-off" $G(E + i\epsilon)$ 
and get (suppressing the Thomas-Fermi term),

\begin{eqnarray}
(-2\pi i)^{-1}  \hbar\omega G(E + i\epsilon) & = & \frac{1}{2} ~ 
\{ {\cal E} + \alpha_q - \frac{1}{2} ( 1 - 
e^{-i\pi ( {\cal E} + \alpha_q + i\epsilon ) } ) ~ \} ~~ \{
\frac{ e^{2\pi i ( {\cal E} + \alpha_q + i\epsilon)}}{ {1 - e^{2\pi i  
({\cal E} + \alpha_q + i\epsilon)}} } \}  + \nonumber \\
& & \frac{1}{2} ~ \{ {\cal E} - \alpha_q + \frac{1}{2} ( 1 - 
e^{-i\pi ( {\cal E} - \alpha_q + i\epsilon ) } ) ~ \} ~~ \{
\frac{ e^{2\pi i ( {\cal E} - \alpha_q + i\epsilon)}}{ {1 - e^{2\pi i 
({\cal E} - \alpha_q + i\epsilon)}} } \}
\end{eqnarray}

Noting that the second group of braces is sum of a geometric series:

\begin{equation}
\frac{ e^{2\pi i ( {\cal E} \pm \alpha_q + i\epsilon)}}{ {1 - e^{2\pi i 
({\cal E} \pm \alpha_q + i\epsilon)}} }  =  
\sum_{k \ge 1} ~ e^{2\pi i k (\frac{E}{\hbar\omega} \pm \alpha_q + i\epsilon)} 
\end{equation}

we see that the propagator {\it{is}} expressed in a form suggestive of a sum over 
periodic orbits. The first group within the braces being the ``amplitude" while 
the second group being contributions from multiple traversals of a basic periodic
orbit. The two terms can be seen to come from families of elliptical orbits going 
clockwise and anti-clockwise sense and that there {\it {no}} term corresponding to 
degenerate ellipses or half orbits. Note that the terms in the ``amplitudes" 
other than ${\cal E} $ are sub-leading relative to ${\cal E} $. This leading part
of the ``amplitudes" can also be seen to come from POT in presence of
continuous families of periodic orbits \cite{littlejohn}. The ``amplitudes" however
are complex. For $\alpha_q = 0 $ there is cancellation of these sub-leading 
terms and one recovers the exact result for the oscillator. It appears that though
the system is integrable, the trace formula does {\it{not}} give exact
propagator. \\

The poles in $G(E)$ come from the two terms and are given by:

\begin{eqnarray}
E_+(n_+) & = & \hbar \omega ( n_+ - \alpha_q ) ~~~~~~ n_+ ~ \ge 2 \\ 
E_-(n_-) & = & \hbar \omega ( n_- + \alpha_q ) ~~~~~~ n_- ~ \ge 1 
\end{eqnarray}

The residues at these poles ( ${\cal E} \pm \alpha_q = n_{\pm}$ ) are $n_{\pm}$
if $n_{\pm}$ is {\it even} and $n_{\pm} - 1$ if $n_{\pm}$ is {\it odd}. These
residues of course give the degeneracy. {\it Note that the sub-leading terms are
important for this.} \\

It is easy to see that the locations of all the poles can be re-expressed in 
the form given by the exact spectrum. The degeneracies at the poles also match 
exactly as expected. \\

(B) We can also write $(-1)^k cos(k\theta) = cos(k(\theta \pm \pi))$.
Proceeding exactly as in the case (A), we can ``read-off" $G(E + i\epsilon)$
and get, \\

\begin{eqnarray}
\left( \frac{-\hbar\omega}{2\pi i} \right) G(E + i\epsilon) & = & \frac{1}{2} 
\{ {\cal E} + \alpha_q \} \{
\frac{ e^{2\pi i ( {\cal E} + \alpha_q + i\epsilon)}}{ {1 - e^{2\pi i  
({\cal E} + \alpha_q + i\epsilon)}} } \} ~ + 
~ \frac{1}{2} \{ {\cal E} - \alpha_q \} \{
\frac{ e^{2\pi i ( {\cal E} - \alpha_q + i\epsilon)}}{ {1 - e^{2\pi i 
({\cal E} - \alpha_q + i\epsilon)}} } \} \nonumber \\
& & - \frac{1}{4} ~ \{
\frac{ e^{\pi i ( {\cal E} -1 + \alpha_q + i\epsilon)}}{ {1 - e^{\pi i 
({\cal E} -1 + \alpha_q + i\epsilon)}} } \} ~~~ + 
~~~ \frac{1}{4} ~ \{
\frac{ e^{\pi i ( {\cal E} + 1 - \alpha_q + i\epsilon)}}{ {1 - e^{\pi i 
({\cal E} + 1 - \alpha_q + i\epsilon)}} } \} \\
& & \nonumber
\end{eqnarray}

These have poles which are subsumed by the poles from the first two terms. 
The overall pole structure and residues of course are exactly same as before as 
they should be. \\

The exponents in the last two terms however do {\it not} look like the action
integral. A reflecting trajectory will not receive a contribution from the
$\alpha_q$ dependent total derivative term. The reflecting trajectories will
also form a single family of trajectories and thus should give only one term
and not two terms. The last two terms can be combined and the propagator may
be re-expressed as, \\

\begin{eqnarray}
\left( \frac{-\hbar\omega}{2\pi i} \right) G(E + i\epsilon) & = & 
\sum_{k \ge 1} ~ \left[ ~ ( \frac{{\cal E} + \alpha_q }{2} ) ~ 
e^{2\pi i k ({\cal E} + \alpha_q + i\epsilon)}  ~ + ~ 
( \frac{{\cal E} - \alpha_q}{2} ) ~ e^{2\pi i k ( {\cal E} - \alpha_q + i\epsilon)}
 \right. \nonumber \\
& &  \left. ~ + ~ ( \frac{-i \ sin(k \pi \alpha)}{2} ) ~ e^{i \pi k ({\cal E}
- 1)} ~ \right] \\
& & \nonumber
\end{eqnarray}

The propagator now does have form indicating a contribution from the half
period orbits including a contribution from a Maslov index due to reflection.
However it has a complex amplitude with a dependence on $\alpha_c$ and the
multiple traversals index $k$. \\

Thus the exact propagator can be expressed in two different forms mimicking
the trace formula with and without the half period trajectories. Although the
phases are as expected, the amplitudes are not. These terms also have
different orders of $\hbar$. \\

The different orders of $\hbar$ seen can be understood in the context of
trace formula as due to {\it families} of periodic trajectories and not due
to any higher order corrections to the SPA. When periodic trajectories come in
continuous families, the number of Gaussian integrations is reduced since the
integration over the family must be done exactly \cite{littlejohn} and this
increases the powers of $\hbar$ in the denominator. The elliptic trajectories
form a three parameter family while the half trajectories form a one
parameter family which explains the difference in the powers of $\hbar$. \\

While the powers of $\hbar$ can be understood, the precise matching of the
amplitudes from the POT does not seem to follow. The appearance of complex
amplitudes may be indicative of the need for complex trajectories and/or
precise matching of semiclassical wave functions across the excluded regions.
This is beyond the scope of present work. \\

The ambiguous ``presence" of half trajectories seems to suggest that one may
{\it either} exclude these altogether but extend the trace formula to deal with the
non-trivial topology due to the excluded regions {\it or} include these
trajectories as additional families but perhaps use complex trajectories.
Needless to say that the structure of half trajectories will be more
complicated for $N > 2$. For getting the semiclassical eigenvalues, the first
alternative may suffice and is already adequate to indicate the presence of
non-exactly known eigenvalues. As there does not appear to be any scope for going
beyond the leading SPA (all intermediate exponents of phases are quadratic), it 
is not clear how one may develop a semiclassical scheme for obtaining the 
non-linear $\alpha_q$ dependence. \\

\section{Summary and conclusions}

In the previous sections we encountered two important features, one related to the
fundamental group and one related to symmetry reduction due to incomplete
vector field. Both were caused by the same source, namely removal of
coincident points implying topologically non-trivial phase space. Both are 
relevant for periodic orbit theory in a general way and a few general remarks
are in order. \\

The stationary phase approximation to the propagator naturally leads to
periodic orbits in the phase space. These orbits may be isolated or come in
continuous families or both. The continuous families may be generated 
by a full group of symmetries or by only a subset of symmetry transformations. Since
each periodic orbit is also a map of $S^1$ to the phase space $\Gamma$, clearly 
every orbit must belong to one and only one homotopy class of the fundamental 
group, $\pi_1(\Gamma)$, of the phase space. If an orbit is a member of a continuous
family then the entire family must belong to a single homotopy class. Note
that a given homotopy class may contain no orbit (solution of equation of
motion) or several isolated orbits and/or several families of orbits. But a
family can not spill over two distinct homotopy classes. In section III we
saw precisely the splitting of a single basic family for oscillator into two
basic families for anyons because of the non-trivial fundamental group. 
Multiple traversals of course belong to different homotopy classes and are
explicitly summed over. Thus a non-trivial $\pi_1$ 
may (but not necessarily) provide an obstruction to a symmetry.
How exactly may such an obstruction manifest itself? For this we have to
consider vector fields generating symmetries. \\

Recall \cite{abraham} that every function on the phase space generates 
{\it {infinitesimal}} symplectic diffeomorphisms (canonical transformations) 
via its corresponding (globally) Hamiltonian vector field. The Lie algebra of 
such vector field is isomorphic to the Possion Bracket (PB) algebra of functions 
on $\Gamma$.  However such infinitesimal transformations exponentiate to give a one
parameter group of transformations {\it {only if}} the vector field is {\it {
complete}} i.e. the integral curves of the vector field  can be extended
so as to have these as a map from the full $R$. Only complete vector fields -
and this is a global statement - give rise to groups of symmetries. (A
corresponding quantum mechanical statement for continuous symmetries is
contained in the Stone's theorem: every one parameter group of unitary
transformations is  generated by a self-adjoint operator and conversely.) 
Since we usually want classical symmetries to be reflected at the quantum
level with observable generators we have to have {\it {groups}} of symmetries
and hence complete vector fields. \\

The criterion of integrability in terms of vanishing PB's guarantees
only the existence of {\it infinitesimal symmetries} which is of course a 
prerequisite. 
An ``infinitesimally integrable" system may thus be: (a) integrable via
action-angle coordinates if the vector fields are complete and the integral
curves are periodic; (b) integrable, but not by action-angle variables if the
vector fields are complete but only a subset of these have periodic integral
curves and (c) partially integrable if only a subset of vector fields are
complete.  The ``pseudo integrability" property of anyons pointed out in ref.
\cite{pseudo} falls in the category (c). As explained in the appendix A apart 
from the Hamiltonian only the total angular momentum has a complete vector 
field and hence is the only quantum number that survives. \\

A non-trivial fundamental group by itself however does not imply possibility of
incomplete vector fields. For, on a compact manifold all vector fields are
complete and it can of course have a non-trivial $\pi_1$. In many cases
constant energy surfaces are compact and one does not have to worry about
incomplete vector fields. For the anyons however removal of the set $\Delta$
makes the constant energy surface noncompact which admits possibility of
incomplete vector fields and corresponding loss of symmetry. \\

In the present work we have seen manifestation of all these features. \\

We also considered application of the periodic orbit theory to this locally 
trivial (integrable) but globally non-trivial system. 
As a by-product, we saw how the exact spectrum for two anyons is reproduced 
by semiclassical methods. We saw that while the dynamical symmetry is reduced 
from $SU(2)$ to $U(1)$, the rank remained the same and hence integrability property 
is preserved. \\

For $N \ge 3$, we reproduced the previously known exact eigenvalues 
\cite{pseudo}. In addition, we obtained further {\it {new}} linearly 
interpolating eigenvalues. In the language of reference \cite{pseudo}, 
these correspond to effect of `relative' dynamics. In this case the symmetry
reduction was drastic, from $OSp(4N, R)$ to $OSp(4,R) \times O(2,R) \times
O(2,R)$. The rank was reduced from $2N$ to $6$, destroying the integrability 
property.\\

We discussed in detail the issue of ``half" trajectories and pointed out their
ambiguous role. Taking the view-point that ``half" trajectories be excluded,
and noting that there does not seem to be any scope for computing higher
order corrections to the semiclassical spectrum with a possible non-linear
dependence on $\alpha_q$, it seems that the non-linearly interpolating
eigenvalues are genuine quantum consequences beyond what semiclassical
analysis could give. Semi-classical analysis is nevertheless sufficient to
indicate the presence of these eigenvalues. \\

\begin{flushleft}                                         
Acknowledgements: 
\end{flushleft}
It is a pleasure to thank M.V.N. Murthy for pointing out to me the puzzling 
aspect of the half orbits, for discussions and critical reading of the manuscript. 
Thanks are also due to Kapil Paranjape for a discussion on fundamental groups. \\
  
\newpage

{\underline {Appendix A :}} {\bf {OSp dynamical symmetry and classification of
trajectories}} \\

In this appendix we collect together some of the well known facts about
the dynamical symmetry of the $n$ dimensional isotropic oscillator. \\

The isotropic oscillator in $n$ dimensions has $OSp(2n, R)$ as the group
of dynamical symmetries. This is a group of $2n\times 2n$ order
matrices, ${\bf g}$, which are both orthogonal and symplectic. Denoting by
$\bar{{\bf g}}$ the transpose of ${\bf g}$, we have the defining equations: 

\begin{equation}
\begin{array}{llclcl}
& \bar{{\bf g}} ~  {\bf g} & ~  = ~ & {\bf I}_{2n} & : & (\mbox{orthogonal}) \nonumber \\
& \bar{{\bf g}}  ~ {\bf {\Omega}}  ~ {\bf g} & ~ = ~ & {\bf {\Omega}} & : &  (\mbox{symplectic}) \nonumber \\
\mbox{and with} & {\bf g} & ~ \approx ~ & {\bf I}_{2n} + \epsilon {\bf T} & & \mbox{generators $T$ satisfy} \\
& \bar{{\bf T}} & ~ = ~ & - {\bf T} & & \nonumber \\
& \bar{{\bf T}} ~ {\bf {\Omega}} & ~ = ~ & - {\bf {\Omega}} ~ {\bf T} & & \nonumber 
\end{array}
\end{equation}

Here ${\bf I}_{2n}$ is the identity matrix of order $2n$ while ${\bf
{\Omega}}$ is a suitable matrix defining the symplectic condition.  This will
be chosen below. \\

The symplectic condition ensures that we have a (linear) canonical
transformation while orthogonality ensures that the Hamiltonian, 
 $ \sum_i (p_i^2 + q_i^2)/2$ , is invariant. It is easy to see that dimension 
of this group is $n^2$ while its rank is $n$. In fact one can show that the 
group $OSp(2n,R)$ is 
isomorphic to the group $U(n,C)$. Making an explicit choice of {\bf
{$\Omega$}} in a block form, one can obtain the block form for {\bf g} using
the symplectic condition. The real block matrices can be combined into 
complex block matrices. The orthogonality condition in terms of real matrices
then translates into the unitarity condition for the complex matrices
thereby proving the group isomorphism. This isomorphism immediately implies
that the ortho-symplectic group acts transitively on the constant energy sphere. 
This is used in the section III.\\

We are interested in finding the action of this group on the phase space via 
canonical transformations (symplectic diffeomorphisms). For notational 
convenience let us group the standard canonical variables together and denote 
them by $\omega_{\mu} ~, \mu = 1,2, ... , 2n,$ the first $n$ being
coordinates and the last $n$ being the momenta. The dual of the symplectic
form has components, $\Omega^{\mu \nu}$ given by the $\mu \nu$ th element of
the matrix ${\bf {\Omega}}$. With this notation, the PB of functions on the
phase space and the infinitesimal canonical transformations are given by,

\begin{eqnarray}
\{ F(\omega) ~,~ G(\omega) \} & ~ = ~ & \Omega^{\mu \nu} ~ \partial_{\mu} F
~ \partial_{\nu} G \nonumber \\
\delta_{\epsilon}\omega^{\mu} & ~ = ~ & \epsilon ~ \Omega^{\mu \nu} ~
\partial_{\nu}F(\omega)
\end{eqnarray}

Functions purely quadratic in ${\boldmath {\omega}}$ generate linear canonical
transformations and are also closed under the PB's and those which leave the
Hamiltonian invariant give the symmetry transformations.  Explicitly,

\begin{equation}
\begin{array}{lclclcl}
F & ~ \equiv ~ & \frac{1}{2} {\boldmath \bar{{\omega}}} ~ {\bf A} ~ {\boldmath {\omega}} 
& ~ , ~ & 
G & ~ \equiv ~ & \frac{1}{2} \bar{{\bf {\omega}}} ~ {\bf B} ~ {\mathbf {\omega}} 
~~~~~ \Rightarrow ~ \nonumber \\
\{ F , G \} & ~ = ~ & \frac{1}{2} \bar{{\bf {\omega}}} ~ 
( {\bf {A \Omega B - B \Omega A}} ) ~ {\bf {\omega}} & & & & \\
\end{array}
\end{equation}

The Hamiltonian corresponds to $ {\bf A} = {\bf I}_{2n} $. The Poisson
bracket of any $G$ with $H$ vanishes provided the matrix {\bf B} commutes
with ${\bf {\Omega}}$. Thus generic matrices defining quadratic functions are
real, symmetric matrices commuting with ${\bf {\Omega}}$. The matrices ${\bf
{\Omega B}}$ are then antisymmetric and provide an isomorphism of quadratic
functions to the generators ${\bf T}$ of the group $OSp(2n, R)$. \\

Action of the one parameter group generated by a function $G$ is found from
the integral curves of the corresponding Hamiltonian vector field. These
curves are defined by the matrix equations, ($ {\bf T} \equiv {\bf {\Omega
B}} $)

\begin{eqnarray}
\frac{d {\mathbf {\omega }}(\sigma) }{d \sigma} ~ = ~ {\bf {\Omega B \omega}} \\
{\bf {\omega}} (\sigma)  ~ = ~ ( e^{\sigma {\bf {\Omega B}}} ) {\bf {\omega}}( 0
)
\end{eqnarray}

It is convenient to choose a particular grouping of the phase space
coordinates and corresponding choice of ${\bf {\Omega}}$.  Firstly let us
put $n = 2N$, relevant for the present context, so that the phase space is 
$4N$ dimensional. Arrange the coordinates and momenta as $x_1, y_1, p_{1x}, 
p_{1y} , .... , x_N, y_N, p_{Nx}, p_{Ny}$ and denote by $\omega _{i}$ the 
coordinates and momenta of the $i^{th}$ particle. The index $i$ now runs 
from 1, ..., N and each $\omega_i$ is a $4 \times 1$ matrix. Correspondingly 
we choose ${\bf {\Omega}}$ as a block diagonal matrix with N blocks and each 
block being $4 \times 4$ matrix ${\bf {\Lambda}}$. We choose,

\begin{equation}
{\bf {\Lambda}}  ~ = ~ \left( \begin{array}{cc}
{\bf 0}_2 & {\bf I}_2 \\ 
{\bf -I}_2 & {\bf 0}_2 
\end{array} \right)  
\end{equation}

One can choose a basis for $OSp(4N, R)$ as follows. Let us denote the
generators as ${\bf T}_i$, $i = 1,2, ... ,N$ and ${\bf T}_{ij}$ with $i < j$.
Each of these are expressed in the block form. The ${\bf T}_i$ are block
diagonal with a nonzero $4 \times 4$ matrix ${\bf u}_i$ as the $i^{th}$
block element. The ${\bf T}_{ij}$ have a matrix ${\bf v}_{ij}$ at the $i^{th}
$ row and $j^{th}$ column ($(ij)^{th}$ block) and ${\bf {-{\bar{v}}}}_{ij}$ 
at the $(ji)^{th}$ block.  In equations, 

\begin{eqnarray}
( {\bf T}_i )_{mn} ~ = ~ {\bf u}_i ~ \delta_{im} ~ \delta_{mn} \nonumber \\
( {\bf T}_{ij} )_{mn} ~ = ~ {\bf v}_{ij} ~ \delta_{im} \delta_{jn} -
~ {\bf {\bar{v}}}_{ij} ~ \delta_{in} \delta_{jm}
\end{eqnarray}

With these definitions it is easy to translate the conditions on the
generators in terms of the $4 \time 4$ matrices {\bf {u , v}} as:

\begin{equation}
{\bf {\bar{u}}}_{i} ~ = ~ - {\bf u}_{i} ~~~~~~~~~~ 
{\bf u}_{ij}{\bf {\Lambda}} ~ = ~ {\bf {\Lambda}}{\bf u}_{ij} ~~ ; ~~~~~~~~~~ 
{\bf v}_{ij}{\bf {\Lambda}} ~ = ~ {\bf {\Lambda}}{\bf v}_{ij} 
\end{equation}

Thus the {\bf u}'s generate $OSp(4, R)$ while the {\bf v}'s are required to
commute with ${\bf {\Lambda}}$. The number of independent {\bf u}'s is 4
while number of independent {\bf v}'s is 8 of which $4$ are `diagonal' and 
$4$ are `off-diagonal'. The dimension of $OSp(4N, R)$  is thus 
$~4N + 8N(N-1)/2~=~4N^2$. These independent matrices can be explicitly 
chosen in terms of $2 \times 2$ Pauli matrices and the identity matrix as:

\begin{eqnarray}
{\bf u}_{(1,2,3,4)} & ~ \sim ~ &  
\left( \begin{array}{cc}
i{\bf {\sigma}}_2 & {\bf 0}_2 \\ 
{\bf 0}_2 & i{\bf {\sigma}}_2 
\end{array} \right) ~,~ 
\left( \begin{array}{cc}
{\bf 0}_2 & {\bf I}_2 \\ 
-{\bf I}_2 & {\bf 0}_2 
\end{array} \right) ~,~ 
\left( \begin{array}{cc}
{\bf 0}_2 & {\bf {\sigma}}_1 \\ 
-{\bf {\sigma}}_1 & {\bf 0}_2 
\end{array} \right) ~,~ 
\left( \begin{array}{cc}
{\bf 0}_2 & {\bf {\sigma}}_3 \\ 
-{\bf {\sigma}}_3 & {\bf 0}_2 
\end{array} \right) ~;~  \nonumber \\
{\bf v}_{(1,2,3,4)} & ~ \sim ~ &  
\left( \begin{array}{cc}
{\bf {I}}_2 & {\bf 0}_2 \\ 
{\bf 0}_2 & {\bf {I}}_2 
\end{array} \right) ~,~ 
\left( \begin{array}{cc}
{\bf {\sigma}}_1 & {\bf 0}_2 \\ 
{\bf 0}_2 & {\bf {\sigma}}_1 
\end{array} \right) ~,~ 
\left( \begin{array}{cc}
i{\bf {\sigma}}_2 & {\bf {0}}_2 \\ 
{\bf {0}}_2 & i{\bf {\sigma}}_2 
\end{array} \right) ~,~ 
\left( \begin{array}{cc}
{\bf {\sigma}}_3 & {\bf {0}}_2 \\ 
{\bf {0}}_2 & {\bf {\sigma}}_3 
\end{array} \right) ~,~  \\
{\bf v}_{(5,6,7,8)} & ~ \sim ~ &  
\left( \begin{array}{cc}
{\bf {0}}_2 & {\bf I}_2 \\ 
-{\bf I}_2 & {\bf {0}}_2 
\end{array} \right) ~,~ 
\left( \begin{array}{cc}
{\bf {0}}_2 & {\bf {\sigma}}_1 \\ 
-{\bf {\sigma}}_1 & {\bf {0}}_2 
\end{array} \right) ~,~ 
\left( \begin{array}{cc}
{\bf {0}}_2 & i{\bf {\sigma}}_2 \\ 
-i{\bf {\sigma}}_2 & {\bf {0}}_2 
\end{array} \right) ~,~ 
\left( \begin{array}{cc}
{\bf {0}}_2 & {\bf {\sigma}}_3 \\ 
-{\bf {\sigma}}_3 & {\bf {0}}_2 
\end{array} \right) ~,~  \nonumber \\
\end{eqnarray}

We are interested in one parameter groups generated by some element {\bf T}
of the Lie algebra. If {\bf T} satisfies: ${\bf T}^2 = -{\bf P}$ with {\bf P}
satisfying ${\bf P}^2 = {\bf P} \ , \ [ {\bf T} , {\bf P} ] = 0$, then it follows
that, 

\begin{equation}
e^{\sigma {\bf T}} \ = \ {\bf I}_{4N} - {\bf P } + {\bf P} \left[ cos(\sigma) + sin(\sigma)
{\bf T} \right] {\bf P}
\end{equation}

For the basis generators ${\bf T}_i$ we have {\bf P} to be block diagonal
with ${\bf I}_4$ as the $i^{th}$ block while for ${\bf T}_{ij}$ we have {\bf
P} to be the block diagonal matrix with ${\bf I}_4$ as the $i^{th}$ and the
$j^{th}$ blocks. Using these, the exponentials of basis generators can be 
evaluated to get the integral curves as, (with obvious notation)

\begin{eqnarray}
{\mathbf {\omega}}(\sigma) & ~ = ~ & 
\left[ ~ {\bf I}^{\prime}_i \ + \ {\bf I}_i \ \left\{ \ cos(\sigma) ~ + ~ sin(\sigma) 
\ {\bf T}_i \ \right\} \ {\bf I}_i ~ \right] {\mathbf {\omega}} (0) ~~~~ \mbox{and,} \nonumber \\
{\mathbf {\omega}}(\sigma) & ~ = ~ & 
\left[ ~ {\bf I}^{\prime}_{ij} \ + \ {\bf I}_{ij} \ \left\{ \ cos(\sigma) ~ + ~  sin(\sigma) 
\ {\bf T}_{ij} \ \right\} \ {\bf I}_{ij} ~ \right] {\mathbf {\omega}} (0)  
\end{eqnarray}

Therefore integral curves of the basis generators are periodic curves. Further,
the ${\bf T}_i$'s affect only the $i^{th}$ particle position and momenta while
the ${\bf T}_{ij}$ mix the $i^{th}$ and $j^{th}$ particles only. \\

We need to study how the angular momenta $J_i$ and $J_{ij}$ vary under the
action of one parameter subgroups. Define the (block) matrices,

\begin{eqnarray}
({\bf L}_i)_{mn} \ & \equiv & \ {\bf L} \delta_{im} \delta_{mn} ~~~~~~~~
{\bf L} \ \equiv \ {\bf v}_7 \nonumber \\
({\bf L}_{ij})_{mn} \ & \equiv & \ {\bf L} \left\{ \ \delta_{mn} ( \delta_{im}
+ \delta_{jm} ) - \delta_{im} \delta_{jn} - \delta_{in} \delta_{jm} \
\right\} 
\end{eqnarray}

In terms of these matrices the angular momenta are given by:

\begin{eqnarray}
J_i \ & = &  \ \frac{1}{2} \ {\bf \bar{\omega} \ L}_i \ {\bf {\omega}} ~~~~~~~
= \ \frac{1}{2} \ {\bf \bar{\omega}}_i \ {\bf L} \ {\bf {\omega}}_i \nonumber
\\
J_{ij} \ & = & \ \frac{1}{2} \ {\bf \bar{\omega} \ L}_{ij} \ {\bf {\omega}} ~~~~~~
= \ \frac{1}{2} \ {\bf \bar{\omega}}_{ij} \ {\bf L} \ {\bf {\omega}}_{ij}
~,~~~~~ {\bf {\omega}}_{ij} \ \equiv \ {\bf {\omega}}_i - {\bf {\omega}}_j .
\end{eqnarray}

Under the group generated by the basis generator, ${\bf T}_i$,  the $J_m$ for 
instance varies as:

\begin{eqnarray}
2J_m(\sigma) & = & 2 J_m + \delta_{im} \left[ \ -sin^2(\sigma) \ \left\{
2 J_i - {\bf {\bar{\omega}_i \ (\bar{u}\ L\ u ) \ \omega_i}} \right\}
\right. \nonumber \\
& & ~~~~~~~~~~~~~~~ \left. + \ sin(\sigma) cos(\sigma) \ \left\{ \ {\bf {\bar{\omega}_i \ ( 
\bar{u} L + L u) \ \omega_i \ } } \right\} \ \right]
\end{eqnarray}

while under the group generated by the basis generator, ${\bf T}_{ij}$, 
the $J_m$ varies as:

\begin{eqnarray}
2J_m(\sigma) & = & 2 J_m + \delta_{im} \left[ \ -sin^2(\sigma) \ \left\{
2 J_i - {\bf {\bar{\omega}_j \ (\bar{v}\ L\ v ) \ \omega_j}} \right\}
\right. \nonumber \\
& & ~~~~~~~~~~~~~~~ \left. + \ 2 sin(\sigma) cos(\sigma) \ \left\{ \ {\bf {\bar{\omega}_i \ ( 
L v ) \ \omega_j \ } } \right\} \ \right] \nonumber \\
& & ~~~~~~~~~ \delta_{jm} \left[ \ -sin^2(\sigma) \ \left\{
2 J_j - {\bf {\bar{\omega}_i \ (v\ L\ \bar{v} ) \ \omega_i}} \right\}
\right. \nonumber \\
& & ~~~~~~~~~~~~~~~ \left. - \ 2 sin(\sigma) cos(\sigma) \ \left\{ \ {\bf {\bar{\omega}_i \ ( 
v L ) \ \omega_j \ } } \right\} \ \right] ~~~~~~
\end{eqnarray}

Similar but more complicated expressions follow for the $J_{mn}(\sigma)$
also. \\

{\underline {Remark:}} For the case of $N = 1$ we have only one block. This 
could be either a single
two dimensional oscillator OR the relative coordinate dynamics of two anyons.
The generators are of course only ${\bf u}$'s. Of these ${\bf u}_2$
corresponds to the Hamiltonian itself while ${\bf u}_1$ generates same rotations
of both $\vec{r}, \vec{p}$. These two matrices commute with {\bf L} while the
remaining two anti-commute with {\bf L}.  This leads to the result quoted in
the section III . \\

To deduce the surviving symmetry group for many anyons we need infinitesimal
variations of $J_{mn}$ for arbitrary generator {\bf T}.  This is easily
derived and is given by,

\begin{equation}
\delta J_{mn} \ = \ \sum_{j} {\bf {\bar{\omega}_{mn}\ L \ ( T}}_{mj} - {\bf T}_{nj} \
) {\bf {\omega}}_j 
\end{equation}

In section III we needed to determine {\bf T} such that $\delta J_{mn}$ is
zero {\it whenever} $J_{mn}$ is zero. That is, at all points in the phase
space where any single $J_{mn}$ is zero, we want its infinitesimal variation
induced by {\bf T} to be zero. \\

Fix a particular $m$ and $n$. Fix ${\bf {\omega}}_{mn}$. We can consider
points with ${\bf {\omega}}_j, \ j \ne m , n$ such that no other $J_{ij}$ is
zero. $\delta J_{mn} \ = \ 0$ then implies ${\bf T}_{mj} \ = \ {\bf T}_{nj} \
\forall \ j \ne m , n$. We can also consider points with different ${\bf
{\omega}}_m , {\bf {\omega}}_n$ keeping ${\bf {\omega}}_{mn}$ fixed and
maintaining all other conditions. This implies ${\bf T}_{mm} - {\bf T}_{nm} +
{\bf T}_{mn} - {\bf T}_{nn} = 0$. Repeating this for all $m \ne n$ fixes the
form of {\bf T} in terms of arbitrary generators of $OSp(4,R)$, {\bf u} ,
{\bf v} as,

\begin{equation}
({\bf T})_{ij} \  =  \ {\bf u} \delta_{ij} \ + \ {\bf v} ( 1 - \delta_{ij} )
~~,~~ \left[ \ {\bf L} \ , \ {\bf u - v}\ \right] \  =  \ 0 
\end{equation}

The choice {\bf u} = {\bf v} corresponds to ${\bf T}_{ij} \ = \ {\bf u} ~ 
\forall \ i, j$.  It follows that {\it all} $J_{mn}$ 's are invariant independent
of their values. It is easy to see that these {\it four} {\bf T}'s
effect transformations of the center-of-mass variables which are insensitive
to the anyonic features. This $OSp(4,R)$ symmetry is thus always present for all 
$N \ge 2$. \\

The choice {\bf v} = 0 implies {\bf T} is block diagonal with the same {\bf
u} on all the diagonal blocks. Further, ${\bf [ L , u ]} = 0$ implies that
{\bf u} must be a linear combination of ${\bf u}_1$ and ${\bf u}_2$ .
These {\it two} {\bf T} 's can be seen to correspond to the total Hamiltonian 
and the total angular momentum. This result is used in the section III to deduce 
that the surviving symmetry for $N$ anyons is $OSp(4,R) \times O(2,R) \times O(2,R)$.

\newpage

{\underline {Appendix B :}} {\bf {Regularised classical dynamics and 
reflecting orbits }} \\

Consider without loss of generality the case of two anyon in the relative 
coordinates. Generic orbits in the configure space, $R^2 - \{\vec{0}\}$,
are of course an ellipse and only a degenerate elliptical orbit attempts to pass 
through the origin. Noting that this system can be thought of as a charged 
particle in presence of a singular (``statistical") magnetic field 
at the origin, one may regularise the magnetic field or the flux to study the 
orbits and obtain the limiting behaviour to deduce ``boundary condition" at the
origin. \\

To do this we imagine the relative Hamiltonian to be that of a particle in a
magnetic field along the z axis. The field is of course to be effectively
confined to a small disc around the origin. Now observe that for an axially
symmetric magnetic field along the z-axis, $\vec{B(r,\theta)} = B(r) \hat{k} $ 
where $r, \theta$ are the usual spherical coordinates in two dimensions,  we may 
write the vector potential as,

\begin{equation}
\vec{A} = A_r \hat{r} + A_{\theta} \hat{\theta} 
\end{equation}
 
This implies,

\begin{equation}
B(r) = \partial_r A_{\theta} + \frac{A_{\theta}}{r} -\frac{1}{r} \partial_{\theta} A_r ,
\end{equation}
 
In a symmetric gauge, the vector potential is independent of the polar angle
and by choosing it to be divergence free one can set the radial component to
zero. \\

The flux $\Phi(R)$, through a disc of radius $R$ is given by,

\begin{equation}
\Phi(R) \ = \ 2 \pi \int_0^R dr r B(r) ~~ = R \oint A_{\theta} d\theta \ = \ 
2 \pi R A_{\theta}(R).
\end{equation}

This implies,

\begin{eqnarray}
A_{\theta}(r) = \frac{\Phi(r)}{2 \pi r} = 
\frac{1}{r} \int_{0}^{r} dr^{\prime} r^{\prime} B(r^{\prime}) 
\end{eqnarray}

The choice, $\Phi(r) \ = \ 2 \pi \alpha ~ \forall \ r > 0$, gives the vector
potential used in the quantum mechanical calculation. It also implies the $r
B(r) \ = \ \alpha \delta(r)$ and this of course is the singular nature of the
magnetic field. Notice that smearing the $\delta(r)$ will not make the
magnetic field non-singular at the origin because of the explicit $1/r$.\\

For a regulated system one wants the magnetic field to be non-singular every
where and effectively confined to a disk of radius $\epsilon$ around the
origin. This requires the vector potential $A_{\theta}$ also to be non-singular and
therefore vanishing at the origin. If $A_{\theta}(r) \ \rightarrow \ c
r^{\beta}$ as $r$ approaches zero then, $B(r) \ \rightarrow \ c(\beta + 1)
r^{\beta -1}$. For a finite, nonzero $B(0)$ we must have $\beta \ = \ 1$ and
therefore $\Phi(r) \ \rightarrow \ 2 \pi c r^2$ near the origin. For $r \ge
\epsilon$ we still retain the flux to be $2 \pi \alpha$. Continuity at $r =
\epsilon$ then gives $c = \alpha / \epsilon^2$. Notice that this limiting
behaviour is fixed by the demand of non-singularity of the fields and as such
must be reflected in any explicit choice for the magnetic field. \\

Thus our regulation involves choosing,

\begin{eqnarray}
\Phi(r) & = & 2 \pi \alpha ~~~~~~~~ \forall ~~ r \ge \epsilon \nonumber \\
& \rightarrow &  2 \pi \alpha \frac{r^2}{\epsilon^2} ~~~~ \mbox{as} ~~ r \ 
\rightarrow 0 
\end{eqnarray}

One could choose a uniform nonzero magnetic field inside the disk as an
explicit choice but it will not be necessary. Since we are interested in the
limiting behaviour of trajectories as $\epsilon$ is taken to zero, the
limiting behaviour of the flux is all that we need. \\

Consider now the orbit equation for $\epsilon$ nonzero. Integrating
the equations of motion once using the two constants, energy $E$ and angular
momentum $\ell$, the orbit equations in $r, \theta$ coordinates become :

\begin{eqnarray} 
\dot{r} ~~~~~ & = &~~~~~\pm \sqrt{2E - r^2 {\dot{\theta}}^2 - r^2 }   \nonumber \\
r^2 \dot{\theta} ~~~~~ & = & ~~~~~ \ell - r A_{\theta}(r) \\
& = & ~~~~~ \ell - \alpha \frac{r^2}{\epsilon^2} ~~~~~~~~~~ 
r ~ < ~ \epsilon \nonumber \nonumber \\
& = & ~~~~~ \ell - \alpha ~~~~~~~~~~~~~ r ~ \ge ~ \epsilon \nonumber 
\end{eqnarray} 

There are three types of orbits possible: those which are fully inside the
disc, those which are fully outside the disc and those which go both inside
and outside the disc. In the limit of $\epsilon$ going to zero, the first type
of orbits are clearly irrelevant. It is easy to see that for the second type
of orbits one must have $\ell \neq \alpha$. These are insensitive to the flux 
in the limit and are thus identical to the orbits of the oscillator.\\

For the last type, an interior turning point is possible
only for $0 \le \ell \le \alpha$. The condition that such an orbit must also
have an exterior turning point limits $\ell$ to $\alpha$. We are interested in 
computing the change in the angular coordinate from the entry into the disk till 
exit from it. Explicit computation shows that the change in the angular coordinate 
goes to zero as $\epsilon$ goes to zero. Thus such an orbit {\it reflects} at 
the origin. Note that these are precisely the radial orbits.  \\

To summarize, a regulated classical modeling for anyons is generically
stipulated by giving the behaviour of the flux near the origin. It amounts to
cutting out a disk of radius $\epsilon$ and filling it up with non-singular
fields. The classical non-radial orbits then are exactly same as those of the 
oscillator except for the replacement $l \rightarrow l -\alpha$. The radial 
orbits reflect at the the origin and are termed {\em half orbits} since their period 
is half of that for the other orbits. 

\end{document}